# Electron-Energy Loss of Ultraviolet Plasmonic Modes in Aluminum Nanodisks

Yujia Yang,[1] Richard G. Hobbs,[1,2] Phillip D. Keathley,[1] Karl K. Berggren[1]

[1]Research Laboratory of Electronics, Massachusetts Institute of Technology, Cambridge, Massachusetts 02139, United States

[2]Centre for Research on Adaptive Nanostructures and Nanodevices (CRANN) & Advanced Materials Bio-Engineering Research Centre (AMBER), School of Chemistry, Trinity College Dublin, Dublin 2, Ireland

**Abstract**

We theoretically investigated electron energy loss spectroscopy (EELS) of ultraviolet surface plasmon modes in aluminum nanodisks. Using full-wave simulations, we studied the impact of diameter on the resonant modes of the nanodisks. We found that the mode behavior can be separately classified for two distinct cases: (1) flat nanodisks where the diameter is much less than the thickness; and (2) thick nanodisks where the diameter is comparable to the thickness. While the multipolar edge modes and breathing modes of flat nanostructures have previously been interpreted using intuitive, analytical models based on surface plasmon polariton (SPP) modes of a thin-film stack, it has been found that the true dispersion relation of the multipolar edge modes deviates significantly from the SPP dispersion relation. Here, we developed a modified intuitive model that uses effective wavelength theory to accurately model this dispersion relation with significantly less computational overhead compared to full-wave electromagnetic simulations. However, for the case of thick nanodisks, this effective wavelength theory breaks down, and such intuitive models are no longer viable. We found that this is because some modes of the thick nanodisks carry a polar (*i.e.* out of the substrate plane, or along the electron beam direction) dependence and cannot be simply categorized as radial breathing modes or angular (azimuthal) multipolar edge modes. This polar dependence leads to radiative losses, motivating the use of simultaneous EELS and cathodoluminescence measurements when experimentally investigating the complex mode behavior of thick nanostructures.

**Introduction**

Electron energy loss spectroscopy (EELS) in a scanning transmission electron microscope (STEM) has recently emerged as an alternative to optical characterization for probing the surface plasmon properties of nanostructures [1–15]. While full-wave electromagnetic calculations can accurately resolve the plasmonic properties of nanostructures, they are computationally intensive. The nanostructures prepared with lithographic techniques for EELS measurement typically possess a flat geometry with their lateral dimensions much larger than the thickness ("flat nanostructures"), motivating the use of intuitive analytical models [4]. These intuitive analytical models commonly used divide the plasmon modes into a combination of multipolar edge modes (modes bound to the periphery of the nanostructure as shown on the left in Figure 1a), and breathing modes (modes confined within a two-dimensional cavity formed by the interface between the nanostructure and the supporting substrate as shown on the right in Figure 1a). While these intuitive analytical models accurately capture the behavior of breathing modes, they fail to describe the experimentally observed dispersion relation of multipolar edge modes. Motivated by experimental results that indicate the modal dispersion relations resemble those of plasmonic waves propagating along the edge of the nanostructures [13,14], in this work we developed a modified intuitive model that uses a two-dimensional cross-section mode solver to characterize the multipolar edge modes without resorting to three-dimensional full-wave calculations. We show that this approach provides both a fast and accurate computational method for simulating the electromagnetic properties of flat nanostructures often studied in EELS experiments, without sacrificing the intuitive appeal of less accurate, fully-analytical approaches.

While flat nanostructures permit the analysis using our modified intuitive model, for the case of "thick nanostructures" where the diameter is comparable to the thickness (see Fig. 1b), such models no longer apply. As the dimension of lithographically defined nanostructures for STEM-EELS measurement approaches the few-nanometer regime [16], it is less clear how the plasmonic modes behave and couple to the exciting electron beam. For such structures, overlapping surface plasmon peaks and the limited energy resolution of electron spectrometers have prevented the experimental study of plasmonic modes in small-diameter, thick nanodisks (for example, see Ref. [16]). Using full-wave simulations, we show that unlike flat nanostructures, for thick nanostructures some modes carry a polar (out of the substrate plane, or along the electron beam direction) dependence. The polar nature of the breathing modes in thick nanostructures creates a net dipole moment in the vertical direction, indicating these modes are bright and can couple to free-space radiation. To investigate these bright modes, we also simulated the cathodoluminescence (CL) to characterize their far-field radiation properties.

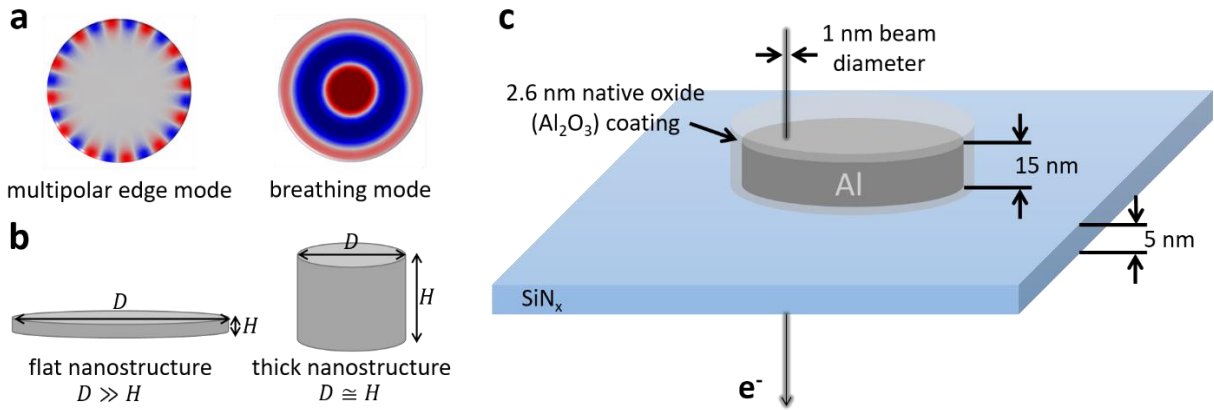

**Figure 1.** Schematic of the aluminum nanodisk structure. **a**, Depiction of the charge distributions for a multipolar edge mode and a breathing mode. **b**, Illustration of a flat nanostructure and a thick nanostructure. **c**, Schematic of the simulation with an electron beam as the electromagnetic excitation. The nanodisk thickness is 15 nm, and is supported by a silicon nitride film with 5 nm thickness. The aluminum core is surrounded by a 2.6-nm-thick native oxide coating. The electron

beam is considered as a linear current with a direction perpendicular to the nanodisk and supporting film.

**Simulated EELS and ultraviolet plasmonic modes of aluminum nanodisks**

The structure under investigation was an isolated aluminum nanodisk supported by a silicon nitride thin film (Figure 1c). This configuration has been commonly used in EELS experiments studying surface plasmons of nanoparticles [3–6,9,10,12–17]. We chose aluminum as the plasmonic material. Aluminum has recently attracted a growing interest as a novel plasmonic metal [16,18–24], as it is cheap, naturally abundant, and CMOS-compatible. Unlike gold or silver, aluminum nanostructures support high energy surface plasmon resonances at visible to ultraviolet and deep-ultraviolet wavelengths. The nanodisk thickness was 15 nm, and its diameter was varied to study the geometry-dependent plasmonic resonances. The sharp edges of the nanodisk were rounded with a 3 nm curvature to avoid singularities and to better represent experimentally fabricated nanostructures. We included a 2.6-nm-thick native oxide (alumina) coating surround the aluminum core [22]. The silicon nitride film thickness was 5 nm. The simulation was performed by a commercial electromagnetic solver *COMSOL Multiphysics*. More details about the modeling can be found in Appendix A.

Figure 2 shows the simulated electron energy loss spectra and plasmonic mode profiles of an aluminum nanodisk with 120 nm diameter. Two configurations with different electron-beam positions were considered: one at the nanodisk edge and the other at the nanodisk center. Multiple EELS peaks corresponding to various plasmonic modes were observed in both configurations (Figure 2a). To differentiate between the two configurations, we named the modes as surface plasmon (SP) modes for electron-beam at the edge surface, and center plasmon (CP) modes for electron-beam at the nanodisk center. We labeled the three lowest modes (SP1 at 2.5 eV, SP2 at 3.4 eV, and SP3 at 4.0 eV) for electron-beam at the edge, and the two lowest modes (CP1 at 4.7 eV and CP2 at 5.7 eV) for electron-beam at the center. Figures 2b and 2c demonstrate the surface normal electric field profiles of the modes for the two electron-beam configurations. For the case with the electron-beam at the edge (Figure 2b), the mode profiles show the three modes are multipolar edge modes: dipole mode (SP1), quadrupole mode (SP2), and hexapole mode (SP3). For the case with the electron-beam at the center (Figure 2c), the mode profiles show the two modes are $1^{st}$ (CP1) and $2^{nd}$ (CP2) order breathing modes. The electron beam excitations of multipolar and breathing modes shown here are consistent with previous reports [3,4,17].

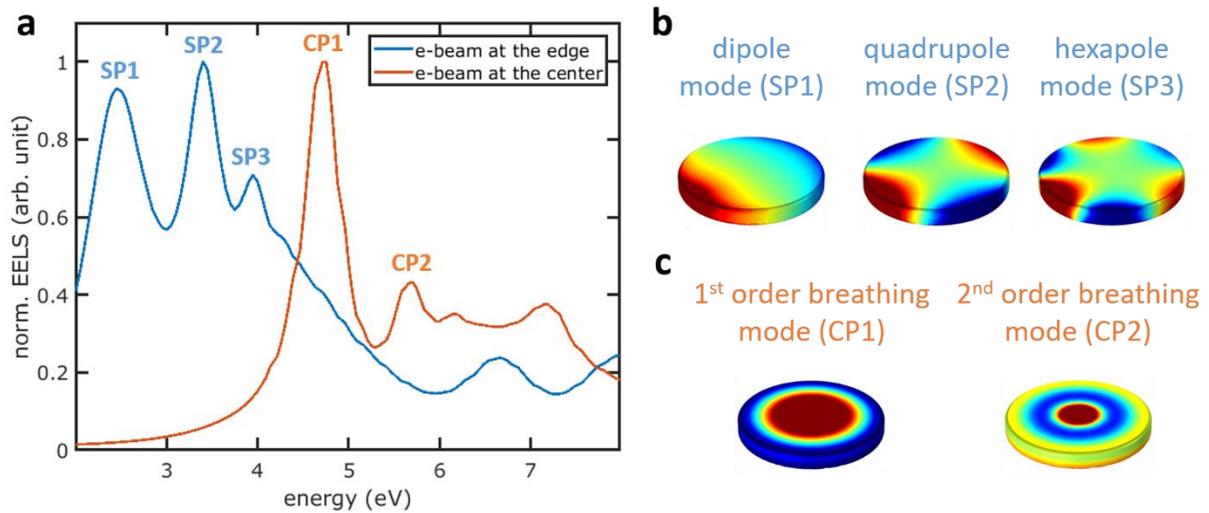

**Figure 2.** Simulated electron energy loss spectra and plasmonic mode profiles of the aluminum nanodisk. **a,** Normalized electron energy loss spectra of an aluminum nanodisk with 120 nm diameter. The electron beam is either at the edge (blue) or at the center (orange) of the nanodisk. The labeled peaks correspond to three lowest modes (SP1, SP2, SP3) for electron-beam at the edge and two lowest modes (CP1, CP2) for electron-beam at the center. **b,** Surface normal electric field

profiles of the three lowest modes for electron-beam at the edge. According to the mode profiles, SP1 is the dipole mode, SP2 is the quadrupole mode, and SP3 is the hexapole mode. **c,** Surface normal electric field profiles of the two lowest modes for electron-beam at the center. According to the mode profiles, CP1 is the 1$^{st}$ order breathing mode, and CP2 is the 2$^{nd}$ order breathing mode. In **b** and **c**, the color scale is saturated to better visualize the mode profiles.

We studied the ultraviolet plasmonic modes of aluminum nanodisks with diameters ranging from 120 nm down to 20 nm. Figure 3 shows the simulated EELS spectra and the plasmonic mode resonant energies for these nanodisks. As expected from the nanoparticle-size-dependency of localized surface plasmon resonance, the multipolar and breathing modes were blue shifted with decreasing nanodisk diameter [22]. For instance, the SP1 mode was shifted from 2.5 eV (corresponding to 496 nm free-space wavelength) for a 120-nm-diameter nanodisk to 5.2 eV (corresponding to 238 nm free-space wavelength) for a 20-nm-diameter nanodisk, and the CP2 mode was shifted from 5.7 eV (corresponding to 218 nm free-space wavelength) for a 120-nm-diameter nanodisk to 6.7 eV (185 nm free-space wavelength) for a 40-nm-diameter nanodisk. By changing the aluminum nanodisk diameter, the plasmonic modes can be tuned from visible to vacuum ultraviolet spectral range.

We note the plasmonic modes for a 20-nm-diameter nanodisk behave unexpectedly. For a 20-nm-diameter nanodisk with the electron-beam at the nanodisk edge (Figure 3a), the second plasmonic mode appears at 6 eV. However, this mode cannot be categorized as the quadrupole mode (SP2), given that the relative intensity of the quadrupole mode (SP2) compared to the dipole mode (SP1) keeps decreasing for a nanodisk diameter changing from 120 nm to 40 nm, but suddenly increases for the 20-nm diameter. By inspecting the mode profile, we found this mode is the 1st order breathing mode (CP1). The identification of this mode is also confirmed by the simulated EELS spectrum for a 20-nm-diameter nanodisk with the electron-beam at the nanodisk center (Figure 3b), in which the first plasmonic mode appears at 6 eV. However, the energy of this mode does not follow the trend of other CP1 modes for the nanodisks with a diameter ranging from 120 nm to 40 nm, as it is red shifted compared to the CP1 mode of a 40-nm-diameter nanodisk. For nanodisks with small diameters, intuition based on flat nanostructures [4] is no longer valid, causing the unexpected mode energy. We will also discuss it later for a 12-nm-diameter nanodisk.

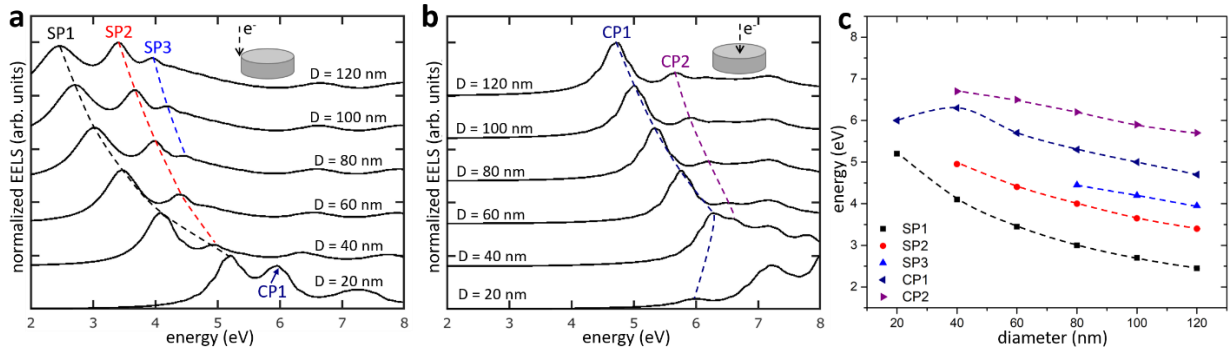

**Figure 3.** Simulated EELS and plasmonic mode resonant energies for aluminum nanodisks with different diameters (20 nm – 120 nm). The normalized EELS spectra are simulated with the electron-beam at the nanodisk edge (a) and at the nanodisk center (b). The spectra for different nanodisk diameters are shifted vertically for clarity. The nanodisk diameters *D* are labeled alongside the spectra. The dashed curves are guides to the eye showing the change of the resonant energies of several plasmonic modes (SP1, SP2, SP3, CP1, CP2) with a varying nanodisk diameter. (c) Plasmonic mode resonant energies extracted from the simulated EELS. Data points with different colors and symbols represent different plasmonic modes (black squares: SP1, red circles: SP2, blue triangles: SP3, navy blue left-pointing triangles: CP1, purple right-pointing triangles: CP2). The dashed curves are guides to the eye.

**Dispersion relations of multipolar and breathing modes**

The dispersion relation of nanoparticle plasmonic modes can be obtained by scaling the modes to surface plasmon polariton (SPP) modes propagating at an extended thin film or at the edge of a thin film [3–5,12,14]. Specifically, the multipolar modes of plasmonic nanoparticles can be considered as surface plasmon edge modes (edge plasmon modes, or EP modes) propagating and resonating at the periphery of the nanoparticles, while the breathing modes can be considered as thin film SPP modes confined in a two-dimensional cavity defined by the nanoparticle geometry. We show the dispersion relation of aluminum nanodisk plasmonic modes in Figure 4. The breathing modes can be considered as the thin film SPP modes confined in the nanodisk cavity, with the surface plasmon wavenumber $k$ satisfying the following relation

$$k_n D = 2n\pi - \phi$$

Here, $n$ is the mode order, $D$ is the nanodisk diameter, and $\phi$ is the nontrivial phase shift upon reflection at the nanodisk boundary [14,25,26]. Figure 4a shows the dispersion relation for the antisymmetric SPP modes of an extended thin film stack consisting of 15 nm aluminum and 5 nm silicon nitride, as well as the dispersion relation of the first and second breathing modes interpreted as nanodisk cavity modes. To fit the breathing modes and SPP mode dispersion relation, a phase shift $\phi = 0.6\pi$ was used.

The dispersion relation for the multipolar edge modes can be calculated considering the surface plasmon edge mode (EP mode) is circulating at the nanodisk periphery:

$$k_n \pi D = 2n\pi$$

Here, we dropped the phase shift term as we argue that no reflection boundary is encountered when the edge mode is circulating at the periphery. The dispersion relation of multipolar modes usually deviates slightly from the dispersion relation of the antisymmetric SPP mode [4,14]. This deviation is caused by the different effective wavelength (and hence wavenumber) of the surface plasmon edge mode (EP mode) compared with the antisymmetric SPP mode, as they are associated with different geometries. The SPP mode is associated with an infinite metal-substrate thin-film stack, while the EP mode is associated with the edge of a semi-infinite metallic film on top of an infinite substrate film. To get a better fitting of the multipolar modes, we numerically calculated the exact dispersion relation of the surface plasmon edge mode (EP mode) via a two-dimensional (2D) mode solver in *COMSOL*. Figure 4b shows the dispersion relation for the surface plasmon edge mode (EP mode), as well as the first, second, and third multipolar modes of the nanodisk. It can be seen that the dispersion relation of the EP mode serves as a better fit for the multipolar modes compared to the dispersion relation of antisymmetric SPP mode. Figure 4c&d shows the electric and magnetic field profiles of the fundamental EP mode supported by a semi-infinite aluminum thin film with 15 nm thickness on an infinite silicon nitride film with 5 nm thickness. The semi-infinite aluminum film on an infinite silicon nitride film forms a plasmonic edge waveguide. This EP mode exists for semi-infinite aluminum films both with and without the alumina coating. The field penetration depth along the horizontal direction is on the order of 10 nm, which is comparable to the nanodisk diameter, suggesting the coupling of charges across the nanodisk causes mode distortions and hence the minor deviation of the multipolar modes dispersion relation from the EP mode dispersion relation. The accurate modeling of the multipolar edge modes as waveguide modes is consistent with previous reports [13,14] in which the modal dispersion relations were experimentally measured from waveguiding nanostructures.

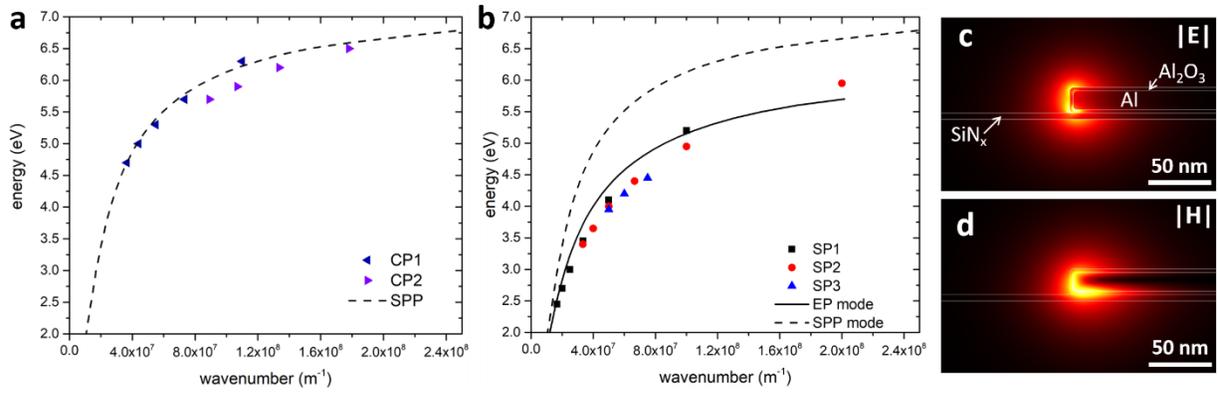

**Figure 4.** Simulated dispersion relation of plasmonic modes. **a,** Dispersion relation of the breathing modes (CP1: navy blue left-pointing triangles, CP2: purple right-pointing triangles). The dashed black curve (SPP) shows the dispersion relation of the surface plasmon polariton (antisymmetric mode) of a thin film stack consisting of 15-nm-thick aluminum and 5-nm-thick silicon nitride. The fitting considers a $0.6\pi$ phase shift upon reflection at the nanodisk boundary. **b,** Dispersion relation of the multipolar modes (SP1: black squares, SP2: red circles, SP3: blue triangles). The solid black curve shows the dispersion relation of the fundamental surface plasmon edge mode (EP mode) propagating along the edge of a semi-infinite 15-nm-thick aluminum film on a 5-nm-thick silicon nitride film. The dashed black curve shows the dispersion relation of the SPP mode as shown in **a**. **c** & **d,** electric ($|\mathbf{E}|$) and magnetic ($|\mathbf{H}|$) field profiles of the fundamental EP mode propagating along the edge. The color scale is saturated to better visualize the mode profiles.

**Simulated EELS and CL of plasmonic modes of a small nanodisk**

We further investigated the surface plasmon modes of a nanodisk with a small diameter (Figure 5). The nanodisk diameter is 12 nm, comparable to its thickness (15 nm). Figure 5a shows the simulated electron energy loss spectra of the nanodisk with different electron-beam positions. Spectra with different colors correspond to different electron-beam positions illustrated in the inset showing the top-view of the nanodisk and the electron-beam positions. The coupling between the electron-beam and the surface plasmon modes depends on the beam position. Therefore, changing the beam position leads to various excitation intensities of the plasmon modes. In Figure 5b, we identified four modes of the nanodisk: first order breathing mode at 4.7 eV, dipole mode at 6 eV, second order breathing mode at 6.4 eV, and a higher order mode at 6.8 eV.

For the nanodisk with a diameter similar to its thickness, the assumption of flat nanostructures is no longer valid, and we found the surface plasmon modes cannot be simply categorized as radial breathing modes or angular (azimuthal) multipolar modes. A polar order is required to describe these modes. The first order breathing mode at 4.7 eV (mode I) can be viewed as a polar mode, with opposite charge polarity at the top and bottom of the nanodisk. This mode carries a net dipole moment in the vertical direction, suggesting it is no longer a dark mode and can be accessed via far-field optical excitation. The bright nature of mode I is further confirmed by the simulated cathodoluminescence (CL) spectra (Figure 5c), showing far-field optical radiation induced by mode I. Due to the polar nature of mode I, it can always be excited by an electron beam in the vertical direction, regardless of the horizontal position of the electron beam. Similar to the nanodisks with a larger diameter, the dipole mode at 6 eV (mode II) can only be excited when the electron beam is away from the nanodisk center, as the mode has an antisymmetric charge distribution in the horizontal plane. Since the mode is bound to the nanodisk edge, the excitation of the dipole mode is stronger when the electron beam is closer to the edge of the nanodisk (x = 2-3 nm). The polar dependence of the plasmon modes is further manifested by mode III at 6.4 eV and mode IV at 6.8 eV. The mode profile of mode III shows a radial dependence of breathing modes in the horizontal plane, together with a polar dependence in the vertical direction. This mode is accessible only when the electron beam is close to the nanodisk center. The mode profile of mode IV shows both an angular dependence similar to the dipole mode in the horizontal plane, and a polar dependence in the vertical direction. This mode is accessible when the electron beam is away from the nanodisk center.

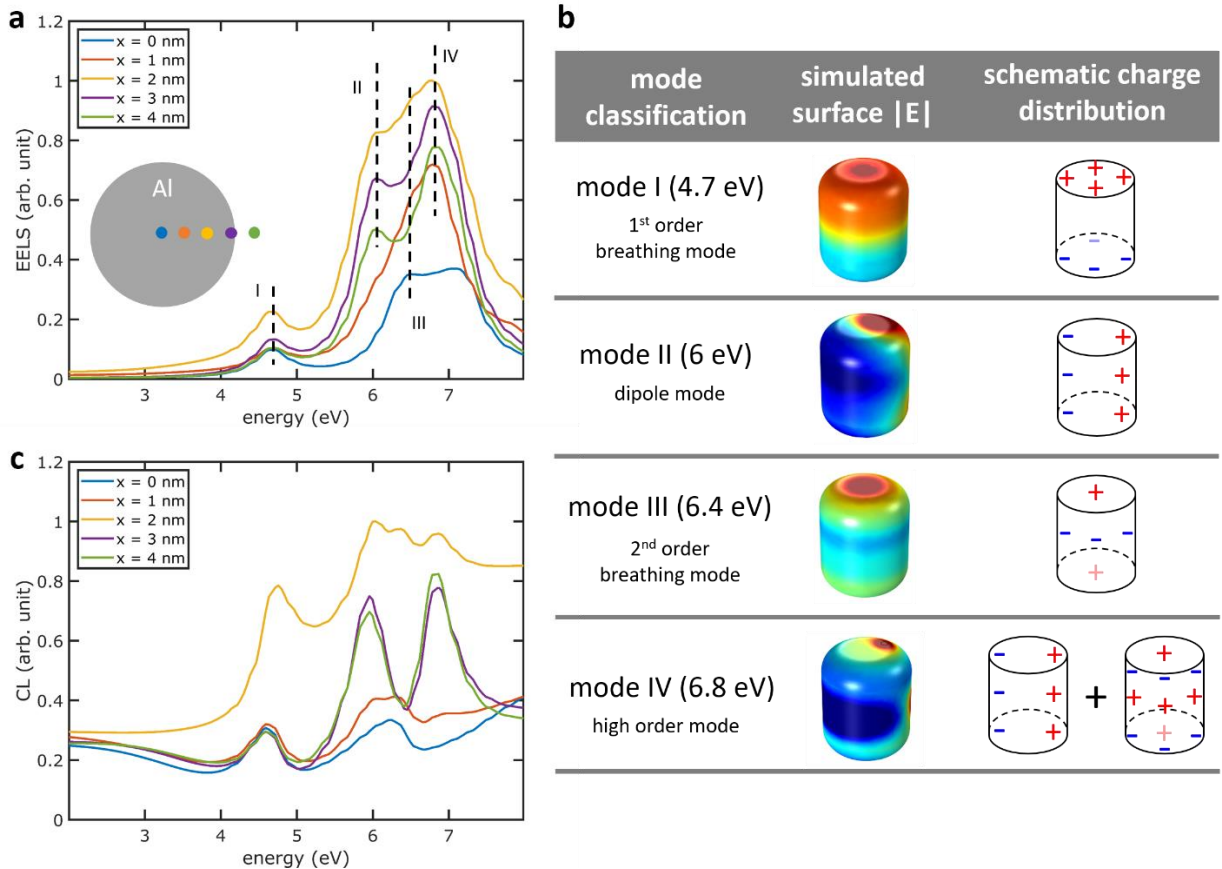

**Figure 5.** Simulated electron energy loss spectra, plasmonic mode profiles, and cathodoluminescence (CL) for an aluminum nanodisk with 12 nm diameter. **a**, Simulated electron energy loss spectra. Spectra with different colors correspond to different electron-beam positions, as illustrated by the inset showing the top-view of the nanodisk and the electron-beam positions. The black dashed lines indicate the mode energies for four plasmonic modes: 1$^{st}$ order breathing

mode (mode I), dipole mode (mode II), 2$^{nd}$ order breathing mode (mode III), and a higher order mode (mode IV). **b**, Table illustrating the mode classification, simulated mode profiles, and the schematic charge distribution profiles. Note some asymmetries in the simulated mode profiles are induced by the placement of the electron beam away from the axis of symmetry of the nanodisk. **c**, Simulated cathodoluminescence spectra. Color encoding is the same as in **a**.

**Conclusion**

In summary, we have theoretically studied the electron energy loss caused by the excitation of surface plasmon modes of aluminum nanodisks with a diameter in the range of ~10 nm to ~100 nm. Our work theoretically demonstrate the plasmonic modes of aluminum nanodisks can be tuned from visible to vacuum ultraviolet spectral range, potentially benefiting UV applications. For nanodisks with a relatively large diameter, the plasmon modes can be characterized as the multipolar modes and the breathing modes. Similar to previous findings, we show that the breathing modes are modeled as cavity modes formed by confinement of the thin film antisymmetric surface plasmon polariton modes in the nanodisk. However, here we model the multipolar modes as ring-resonating modes bound to the nanodisk edge, with the dispersion relation accurately reproduced from a computationally less demanding 2D mode solver. For nanodisks with a diameter comparable to the thickness, we show that the plasmon modes possess a polar nature. This polar nature makes these modes bright, potentially accessible via far-field optical excitation and detection, which previously was only possible for large disks [17]. This suggests that experiment investigating such small-diameter, thick nanostructures would benefit from measurements of the cathodoluminescence, ideally with polarization sensitivity.

**Appendix A: Numerical modeling of EELS in COMSOL**

The electron beam was considered as a linear current induced by an electron moving in the vertical **z** direction with a constant energy. This assumption is valid as the energy loss (~ eV) is much less than the electron energy (100 keV in our simulation). The theoretical treatment was previously reported [8,27] and we give a brief outline here. The spectral current density and electron energy loss probability can be expressed as

$$\boldsymbol{j}(z,\omega) = -\mathrm{e}\hat{\mathbf{z}}\delta[\boldsymbol{r}_t - \boldsymbol{R_0}]e^{\mathrm{i}\omega z/v}$$

$$\Gamma_{\mathrm{EELS}}(\omega) = \frac{v\mathrm{e}}{2\pi\hbar\omega}\int dz\, \mathrm{Re}[e^{-\mathrm{i}\omega z/v}\hat{\mathbf{z}} \cdot \boldsymbol{E}_{\mathrm{in}}(z,\omega)]$$

Here, $\boldsymbol{j}(z,\omega)$ is the current density as a function of position and (angular) frequency, e is the electron charge, $\hbar$ is the reduced Planck constant, $v$ is the scalar velocity of the electron, $\boldsymbol{r}_t = (x,y)$ is the transverse position, $\boldsymbol{R_0} = (x_0, y_0)$ describes the transverse position of the electron beam, and $\boldsymbol{E}_{\mathrm{in}}(z,\omega)$ is the electric field induced by the linear current.

The induced field was calculated by a finite-element electromagnetic solver (*COMSOL Multiphysics*) using the linear current excitation. The excitation electron beam was modeled by a long cylinder carrying the linear current. The cylinder diameter was 1 nm, representing a finite beam spot size. For the numerical simulation results shown in this paper, the electron beam current density was assumed to be constant in the transverse (x, y) plane of the cylinder. Instead of an infinitely narrow line, using a finite-diameter cylinder avoids singularities in the calculation, improves the meshing quality, and allows for the implementation of arbitrary current distribution in the transverse plane. The optical properties for aluminum were taken from Rakić [28] with linear interpolation. The refractive index of alumina and silicon nitride were fixed at 1.88 [29] and 2.4 [30], respectively, as the dispersion of both materials were negligible within the spectral range of interest. Small variations in the dielectric refractive indices could lead to a small shift in the plasmonic mode energy, but would not affect the mode profile. The full calculation domain was encapsulated in a 50-nm-thick spherical perfect matched layer (PML) to absorb outgoing electromagnetic waves without undesired reflections at the domain boundary. The model setup also allowed for the simulation of cathodoluminescence (CL), in which the outgoing power at the simulation domain boundary was calculated. Due to the many different length scales in the model (electron beam diameter of 1 nm, Al nanostructure on the order of 10 nm, calculation domain on the order of 100 nm), entities in the model were meshed adaptively with highly nonuniform tetrahedral discretization elements to ensure a high quality meshing. The EELS was simulated in the energy range of 2 eV to 8 eV with a 0.05 eV energy step. Maxwell's equations were solved in the frequency domain using the multifrontal massively parallel sparse (MUMPS) direct solver [31].

**Acknowledgements**

We would like to acknowledge partial support from Gordon and Betty Moore Foundation. This work was also supported by MIT Center for Excitonics under Award Number DE-SC0001088. We would like to thank Marco Colangelo and Dr. Reza Baghdadi for helpful discussions on the manuscript.